\documentclass[conference]{IEEEtran}

\usepackage{cite}

\usepackage{subfigure}

\usepackage{url}
\usepackage{verbatim,epsf}
\usepackage[caption=false,font=footnotesize]{subfig}

\usepackage{tabularx}
\usepackage{balance} 
\usepackage{amssymb}
\usepackage{amsmath}
\usepackage{graphicx}
\usepackage{epsfig}

\usepackage{comment} 
\usepackage{amsfonts} 
\usepackage{amsmath} 
\usepackage{rotating}

\usepackage{amssymb}
\usepackage{amsmath}
\usepackage{amsfonts}

\usepackage{algorithm2e}

\begin{document}
%
\title{Design and Analysis of Coalitions in Data Swarming Systems}

\author{\IEEEauthorblockN{Honggang Zhang}
\IEEEauthorblockA{Math and Computer Science Department\\
Suffolk University, Boston, MA 02108\\
Email: hzhang@suffolk.edu}
\and
\IEEEauthorblockN{Sudarshan Vasudevan}
\IEEEauthorblockA{Alcatel-Lucent Bell Labs\\
Murray Hill, NJ 07974 \\
Email: sudarshan.vasudevan@alcatel-lucent.com}
}

%

\maketitle

\begin{abstract}
We design and analyze a mechanism for forming coalitions of 
peers in a data swarming system where peers have heterogeneous upload capacities.
A coalition is a set of peers that
explicitly cooperate with other peers inside the coalition via {\em
choking}, {\em data replication}, and {\em capacity allocation} strategies. 
Further, each peer interacts with other peers outside its coalition via potentially distinct
choking, data replication, and capacity allocation strategies. 
Following on our preliminary work in \cite{zhang11icnp} that
demonstrated significant performance benefits of coalitions, 
we present here a comprehensive analysis of the choking and data
replication strategies for coalitions. 

We first develop an analytical model to understand a simple random 
choking strategy as a within-coalition strategy
and show that it accurately predicts a coalition's performance.
Our analysis formally shows that the random choking strategy can help
a coalition achieve near-optimal performance by optimally choosing the
re-choking interval lengths and the number unchoke slots. 
Further, our analytical model can be easily adapted to model 
a BitTorrent-like swarm. 
We also introduce a simple data replication strategy which
significantly improves data availability within a coalition as
compared to the rarest-first piece replication strategy employed in
BitTorrent systems. 
We further propose a {\em cooperation-aware better response strategy}
that achieves convergence of the dynamic coalition 
formation process when peers freely join or leave any
coalition. 
Finally, using extensive simulations, we demonstrate improvements in
the performance of a swarming system due to coalition formation.
\end{abstract}

%
\IEEEpeerreviewmaketitle

\section{Introduction}
There have been many recent studies on BitTorrent-like swarming systems, 
mainly via modeling, measurement, and simulation (see for example, \cite{qiu04bt,
misra09bt.model, lui06bt, legout@rarest, msr05bt, levin08auction}).
However, there is still a lack of significant understanding of 
cooperative behavior and its impact on swarming 
systems, except a very few studies viz.~\cite{rafit09bt,misra10,zhang11icnp}.
Following on our initial study in \cite{zhang11icnp}, we formally investigate
cooperative peer behavior in swarming systems through analytical
modeling, design and extensive simulations.

Similar to most existing works (e.g.,
\cite{menasche10sustainability}), we consider a swarming system where there is a content
publisher (i.e. an initial {\em seed}) that never leaves the system and serves a file
(divided into a set of {\em pieces}) to a heterogeneous population of
peers with different upload capacities.
Besides downloading from the publisher, peers also exchange pieces of the shared
file among themselves. 
Since a basic functionality of a data swarming system is to let users
download data, we investigate whether explicit cooperation among 
a group of peers can significantly reduce their file download completion time.
Such a group is referred to as a \textit{coalition}
\cite{zhang11icnp}. Each peer in a coalition cooperates
with other peers inside the coalition via {\em choking}, {\em data
replication}, and {\em capacity allocation} strategies. 
Further, each peer interacts with other peers outside its coalition via potentially distinct
choking, data replication, and capacity allocation strategies. 
Our initial study \cite{zhang11icnp} demonstrates that a coalition
of peers not only significantly reduces the individual peer download
completion times, but also yields performance benefits to the whole swarm. 
As mentioned in \cite{zhang11icnp}, the notion of coalitions differs
from that of clusters studied in
\cite{legout06clustering}, which are formed as a consequence of the 
selfish nature of peers and the Tit-for-Tat strategy. Unlike
a coalition, the lack of cooperation between peers in a cluster can 
degrade the performance of the peers in the cluster~\cite{misra09bt.model}. 
Another related but different notion is that of a buddy group proposed
in \cite{rafit09bt}. However, we have shown in \cite{zhang11icnp} that our
coalition design significantly outperforms the choking strategy adopted by peers in a buddy group.

The present work differs from our initial work~\cite{zhang11icnp} in
the following important ways:
\begin{enumerate}

\item We present an analytical model for investigating the random
choking strategy used by peers in a coalition.  Our model takes into
consideration that a peer may concurrently 
download multiple distinct pieces, which is not considered in \cite{zhang11icnp}.
We also explicitly model the impact of re-choking interval lengths (i.e., the duration of time elapsed
before a peer decides on a new set of peers to unchoke) and the number
of unchoke slots on the coalition performance, whereas
in \cite{zhang11icnp}, the impact of these two key parameters is only observed via simulations. 

\item We then use the analytical model to optimally design a
coalition. In particular, our model yields optimal values of the
re-choking interval length and the number of unchoke slots in each
re-choke interval. Such a model-based optimal design is an important
distinction of this work from~\cite{zhang11icnp}.

\item In order to improve data availability, we introduce a {\em
Peer-balance Rarest-first Piece Selection strategy} for data
replication in a coalition. Data availability has not been studied in
\cite{zhang11icnp}. 

\item Using extensive simulations on a data set of peer upload capacities collected from real-world
swarming systems, we explore the impact of forming coalitions on the overall
performance of a swarm, and investigate whether coalitions can be
dynamically reached in practice, as peers enter and leave the system.
In \cite{zhang11icnp}, simulations are only conducted on synthetic
data sets of two capacity classes of peers.

\end{enumerate}

\subsection{Main Results}
We make the following important contributions in this paper.
\begin{enumerate}
\item We introduce a detailed analytical model of a coalition in data
swarming systems. This analytical model can be easily
adapted to study different choking strategies for coalitions, and even
to a general BitTorrent-like swarm.
Using our model, we analyze the 
impact of the length of re-choking intervals  and
the number of unchoke slots on the system performance. Further, these two
parameters can be optimally chosen based on our model
so as to minimize the average download completion time. 

\item Using our analytical model, we observe that a coalition of peers adopting a simple
random choking strategy as the within-coalition choking strategy exhibits
near-optimal performance. This is particularly 
appealing since (i) the random choking strategy is simple and easily
implementable in a distributed fashion by peers in a coalition, and
(ii) finding an optimal choking strategy appears infeasible. 

\item We further propose a data replication strategy and show that it 
significantly outperforms the conventional rarest-first strategy
in terms of data availability in a coalition. 

\item Using extensive simulations with a real-world data set of peer upload capacities,
we show that coalitions improve
the overall performance of a swarm if the
majority of peers form a coalition. 
Furthermore, we propose an improved {\em cooperation-aware better
response strategy} (from the scheme proposed in \cite{zhang11icnp})
that achieves convergence of the coalition size (i.e., number
of peers in the coalition), 
even when peers are allowed to freely join or leave any coalition.
\end{enumerate}

\noindent \textbf{Related work.}
This paper follows on our initial study on coalitions in \cite{zhang11icnp}.
Modeling the swarm as a sequence of download stages or stations first appeared in 
Menasche \textit{et al} \cite{menasche10sustainability}. 
Tian  \textit{et al} \cite{tian06infocom} study peer distribution as a function of
the fraction of downloaded file.  
The models in \cite{misra09bt.model}\cite{lui06bt} also study the steady state of a swarming system. 
\cite{ezo09mintime} attempts to minimize average finish time
in P2P networks, but it assumes that the shared
file is broken into infinitesimally small pieces such that there is no forwarding delay.
Rafit \textit{et al} \cite{rafit09bt} propose a buddy protocol for
peers to form buddy groups (similar to coalitions). 
Basics of dynamic coalition formation and the cooperative game theory framework 
can be found in \cite{han09coalition.survey}\cite{tone02dynamic.coalition}\cite{madiman08}\cite{saad08}.
Misra \textit{et al} \cite{misra10} studies cooperation in peer-assisted services, but their model is not
applicable to the swarming systems of interest in this paper.

\subsection{Organization of the Paper}
The rest of the paper is organized as follows. We outline the overall coalition design in Section \ref{sec.overview}.
In Section \ref{sec.choking}, we introduce our analytical model that
accurately predicts the file download time of a coalition
and also yields the optimal parameter settings.
In Section \ref{sec.replication}, we present our data replication
strategy among peers in a coalition, and show that it improves data availability significantly.  
In Section \ref{sec.form.impact}, using simulations we demonstrate 
that coalitions improve data swarming performance as a whole. Finally,
we conclude in Section \ref{sec.con}.

\section{Preliminaries: Overview of Coalition Design} \label{sec.overview}
In this section, we briefly review the overall system design of a coalition (introduced in  \cite{zhang11icnp}).
Throughout this paper, a coalition is defined to be a set of peers that
cooperate with each other within the set and interact with peers not in the set, both 
according to a {\em choking strategy},  a {\em piece selection strategy}, and an {\em upload capacity allocation strategy}.
Different from \cite{zhang11icnp}, we allow peers in a coalition to have different upload capacities. 
As for a coalition, we ideally would like to design the choking, capacity allocation
and piece selection strategies so as to minimize 
the average file download completion time of the peers in a coalition. 
However, our focus in this paper is primarily on designing an
efficient choking and piece selection strategies.
Recall that the upload capacity allocation strategy specifies how a
peer determines the fraction of its upload
capacity that is
allocated to each of its downloaders. As in \cite{zhang11icnp}, we assume that a peer does not
allocate capacity to peers outside its coalition, and equally splits
its upload capacity among all of its downloaders in the same coalition. 
A detailed study of more general capacity allocation strategies is a
topic for future work. 

For completeness, we describe preliminaries introduced
in~\cite{zhang11icnp}. We also re-state our random choking strategy
for a coalition, which will be analyzed in detail in subsequent sections. 

\subsection{Choking Strategy}
In BitTorrent-like swarming systems,
a peer unchokes a fixed number of
peers at points in time spaced $\delta_t$ units apart. This interval
$\delta_t$ is referred to as the {\em re-choking interval}. 
The choking strategy of a peer in a coalition determines the set of peers that this
peer unchokes, and consists of two rules:
one for dealing with peers in the same coalition, referred to as
the {\em within-coalition} choking strategy; and the other for dealing with 
peers outside the coalition.
The within-coalition choking strategy proposed in \cite{zhang11icnp} is as follows. Each peer in a
coalition uniformly at random unchokes $k$ other peers in the same
coalition every $\delta_t$ time units. We refer to this strategy as
the {\em random choking strategy}.
In our previous work \cite{zhang11icnp}, the values of $\delta_t$ and
$k$ were empirically chosen to minimize the average download
completion time for a coalition. 
In this paper, we give a detailed analytical model 
that yields the optimal values for $\delta_t$ and $k$. 
A basic requirement is that the values of $k$ and $\delta_t$ should be chosen 
so that a peer can upload at least a complete piece
to each of its data receivers. 
Regarding the rule to deal with peers outside a coalition, 
for simplicity, we assume in this paper that each peer chokes all other peers outside its
coalition. Note that a peer in a coalition can still receive data from
peers outside the coalition when the peer is optimistically unchoked by other peers. 
Studying different strategies for interacting with peers outside a coalition is a topic for future
investigation. 

We now proceed to describe our analytical model of choking strategy for coalitions.

\section{Model and design of choking strategy} \label{sec.choking}

We now introduce an analytical model that enables analysis of the random choking strategy. This model differs from the one in
\cite{zhang11icnp} in that this model explicitly models 
the impact of re-choking interval $\delta_t$ and the number of unchoke
slots $k$ on the performance of a coalition. As we will see, our model
yields the optimal values of
$\delta_t$ and $k$, which were observed only via simulations in our previous work \cite{zhang11icnp}.

\subsection{Assumptions and Steady State Representation of a Swarm}\label{sec.steady.rep}
As in \cite{zhang11icnp},  we assume that peers arrive to the system according to a Poisson process
with rate $\lambda$, and each peer has an upload capacity of $u_p$ \footnote{Note that $u_p$ can also be thought
of as the average upload capacity of all peers with heterogeneous capacities.}.
We use the random variable (r.v.) $N$ to denote the total
number of peers in the system. We assume that all peers are interested in the same file, which
is divided into $B$ pieces. There is an initial seed (or content
publisher) with upload capacity $u_s$ in the system, and it never
leaves the system. On the other hand, each peer 
leaves the system immediately after it has received all the pieces of
the file. 
The data swarm can be modeled as a queuing system with $B+1$
M/G/$\infty$ queues. The $i$-th queue (with $i=0, 1, 2, ...B-1, B$), also referred to as
{\em download stage} $i$, consists of
peers in possession of exactly $i$ pieces. Let r.v. $N_i$
denote the number of peers in the $i$-th queue and $\bar{N}_i$ its expectation. Peers in the $i$-th
queue jump in the $j$-th queue with rate $\beta_{ij}$. In steady state, $d\bar{N}_i(t)/dt = 0, \forall i=0, 1, 2, ..., B-1$,
i.e., the arrival and the departure rates of peers for a given queue 
are identical. 

Let $\gamma_{ij}=\beta_{ij}/\bar{N}_i$. In other words, $\gamma_{ij}$
denotes the transition rate of an arbitrary peer in $i$-th
queue to $j$-th queue. The reason for introducing $\gamma_{ij}$ is
because $\beta_{ij}$ depends on the downloading rates of individual
peers in the $i$-th queue. In other words, once we know the downloading
rates, we can easily calculate $\beta_{ij}$. Further, it is intuitive
to relate an individual peer's download rate to 
coalition parameters such as $\delta_t$ or $k$, as  
will become clear later in the section.  To calculate $\gamma_{ij}$,
we need to compute downloading rate of a peer and the fraction of time
it actively downloads data in each queue.

The arrival rate of peers to queue $i$ (where $i=1, ..., B$) is given by
\begin{equation}
\lambda_i =   \sum_{\ell=0}^{i-1} \beta_{\ell i} = \sum_{\ell=0}^{i-1} \gamma_{\ell i} \bar{N}_\ell
\end{equation}
and the departure rate of peers from queue $i$ (where $i=0, 1, ..., B-1$) is given by 
\begin{equation}
\mu_i = \sum_{\ell=i+1}^{B} \beta_{i \ell} =\bar{N}_i  \sum_{\ell=i+1}^{B} \gamma_{i \ell}
\end{equation}
Note that for queue $0$, we have $\lambda_0=\lambda$, which is the
arrival rate of peers to the coalition. Similarly, for queue $B$, we
have $\mu_B=\infty$, as we assume that peers immediately leave the system once finish downloading. 

In steady state, the arrival rate of peers to queue $i$ is equal to
the departure rate of peers leaving from queue $i$. Thus,  
\begin{equation}
\sum_{\ell=0}^{i-1} \gamma_{\ell i} \bar{N}_\ell = \bar{N}_i \sum_{\ell=i+1}^{B-1} \gamma_{i \ell}
\label{eqn:rate}
\end{equation}
In the following, we model the transition rates $\gamma_{i,j}$ and $\bar{N}_i$ as functions of $\delta_t$ and $k$.
As shown later, solving the set of equations given in (\ref{eqn:rate})
yields the optimal values of $\delta_t$ and $k$ that minimize the expected download completion time.
 
For simplicity, we use the following notational
simplification in the remainder of this section. We use $P(X)$ to
represent $P(X=1)$, where $X$ is an indicator r.v. 

\subsection{Re-choking Interval $\delta_t$} \label{sec.delta_t}
Let $p_i$ and $p_j$ denote arbitrary peers in the $i$-th and the
$j$-th queues respectively. Suppose that $p_j$ randomly unchokes $k$
other peers in each re-choking interval $\delta_t$.
Let $\bar{U}_j$ denote the average per-connection upload rate of $p_j$. Then, it takes $1/\bar{U}_j$ seconds
for $p_j$ to completely send one piece to each of its downloaders. Suppose that $p_i$ is unchoked
by $p_j$ during a re-choking interval $\delta_t$. During this re-choking
interval, $p_j$ continues to download data from other peers and may
transit to another queue upon downloading one or more pieces. Let
$\bar{D}_j$ ($=\sum_{\ell=j+1}^B \gamma_{j \ell} $) denote its transition rate out of queue $j$. Then the effective time interval that 
$p_j$ unchokes $p_i$ (while $p_j$ is still in queue $j$) is given by $\min(\delta_t, 1/\bar{D}_j)$.
We can divide this effective time interval into $\varphi_j$ time
slots, each long enough for $p_j$ to completely upload a piece to a
downloader. More formally, $\varphi_j= \max \big(1, \min (\delta_t, 1/\bar{D}_j) / (1/\bar{U}_j) \big)$.

\subsection{$p_j$'s expected per-connection upload rate $\overline{U}_j$} \label{sec.uj}
Let $B_{ij}$ be an indicator random variable that is set to 1 when $p_i$ is ``interested'' in 
$p_j$ in steady state i.e. $p_j$ has one or more pieces which $p_i$
does not. Assume that $p_j$ assigns one of the $\varphi_j$ slots to
$p_i$ uniformly at random upon unchoking $p_i$.
Let $B_{ij}^{\ell}$ be an indicator random variable that is set to 1 when $p_i$ is interested in 
$p_j$ if $p_i$ is in $\ell$-th slot unchoked by $p_j$, where $\ell=1,...,\varphi_j$.
Note that $P(B_{ij}^{\ell})$ is a conditional probability. It follows that 
\begin{equation}
P(B_{ij}) = \bigg( \sum_{\ell=1}^{\varphi_j} P(B_{ij}^{\ell} ) \bigg) / \varphi_j 
\label{eqn.pb}
\end{equation}
In Appendix \ref{sec.pb}, we describe the calculation of $P(B_{ij}^{\ell})$.

In \cite{zhang11icnp}, we assumed that in steady state,  
the data pieces that have been downloaded both by $p_i$ and $p_j$ 
are chosen from $B$ pieces independently and uniformly at random. In
particular, we obtain
\begin{equation}
P(B_{ij})= \left\{
\begin{array}{ll}
1- \binom{B-j}{i-j} /\binom{B}{i},  & \quad \text{if } i \ge j,\\
1 , & \quad \text{if } i < j.
\end{array} \right.
\label{eqn.pb.icnp}
\end{equation}
However, in our current model, (\ref{eqn.pb.icnp}) is true only when
$p_i$ is unchoked by $p_j$ and is assigned the first slot. Since $p_i$
can be assigned any of the $\varphi_j$ slots in a re-choking interval, 
(\ref{eqn.pb}) represents a more accurate calculation of $P(B_{ij})$, 
as compared to (\ref{eqn.pb.icnp}).

We assume that all $k$ out-connections of $p_j$ are assigned to other peers
independently, and any out-connection is assigned uniformly at random to all other peers in the system.
Let $\eta_j$ denote the probability that an out-connection of $p_j$ is
{\em active}, i.e., there is ongoing data transmission over the connection. Note that
an out-connection may not be active if the unchoked peer is not
interested in the data possessed by $p_j$. Then,
\begin{eqnarray}
\eta_j &=& \sum_{i=0}^{B-1} \bar{N}_i \cdot P(B_{ij})  / \bigg(\sum_{\ell=0}^{B-1} \bar{N}_\ell \bigg)
\label{eqn.eta}
\end{eqnarray}
We finally obtain the expected per-connection upload rate as
\begin{eqnarray}
\overline{U}_j &=& \sum_{w=1}^k  \frac{u_p}{w} \binom{k}{w} \eta_j^w (1-\eta_j)^{k-w}  
\label{eqn:u_j_coal}
\end{eqnarray}

\subsection{The data transfer connection from $p_j$ to $p_i$. } \label{sec.conn}
Let $S_{ij}$ be an indicator random variable that is set to
$1$ if the connection from $p_j$ queue to $p_i$ is active. Per the
random choking strategy, each peer in
the coalition unchokes another peer in the coalition chosen uniformly
at random. 

Let $A_{ij}$ be an indicator random variable that is set to
$1$ if $p_j$ unchokes $p_i$ in an rechoking interval $\delta_t$. It
follows that
\begin{equation}
P(A_{ij})=\sum_{w=1}^k (1-p)^{w-1} p  
\end{equation}
where $p = 1/\sum_{\ell=0}^{B-1} \bar{N}_\ell$.

Finally,  
$$P(S_{ij}) = P(A_{ij}) P(B_{ij})$$
 
\subsection{Transition Rates $\gamma_{ij}$}\label{sec.transition.rates}
Consider an arbitrary peer $p_i$ in queue $i$. Depending on the number
of active download connections of $p_i$, it can transit to any queue
$j$, where $i < j \leq B$. Let $q_i^\ell$ denote the probability that $p_i$
has $\ell$ active download connections (each of which is downloading a
distinct piece), and let $D_i^\ell$ denote the download rate per piece.
Assuming that all $\ell$ connections complete their piece transfers at about the same time, then
$p_i$ will jump to queue $i+\ell$ with transition rate $q_i^\ell D_i^\ell$. The exact calculation of $q_i^\ell$ and $D_i^\ell$
is computationally very expensive, as it involves an exponential number of combinatorial terms. 
For instance, consider the case when $\ell=3$. We need to consider all possible 
combinations of queues that originate these $3$ active connections. 
As an example, the probability of the event that the connections are from 
queues $2, 3, 4$ is given by
$P(S_{i,2})P(S_{i,3})P(S_{i,4})$. Further, the active connections 
have data rates $\bar{U}_2, \bar{U}_3, \bar{U}_4$. Since the exact
computation of download rates of peers is not practical, we
approximate the transition rates as follows. 

Let $\bar{N}_{up}$ denote the expected number of peers uploading data
in the system in steady state. Then, 
\begin{equation}
\bar{N}_{up}=\sum_{j=1}^{B-1} \bar{N}_j
\end{equation}
We exclude peers in queue $0$ from the calculation of $\bar{N}_{up}$,
since these peers do not have a complete piece yet.

Recall that $P(S_{ij})$ denotes the probability that peer $p_i$ has an
active in-connection from $p_j$, an arbitrary peer in
queue $j$, and is a function of $i$ and $j$. We approximate the system
by assuming that all peers in various queues
have identical probabilities of their upload-connections being active
when attempting data transfer to $p_i$. Let $P(S_i)$ denote this
probability, which is given by
\begin{equation}
P(S_i) = \bigg( \sum_{j=1}^{B-1} \bar{N}_j P(S_{ij}) \bigg) / \bar{N}_{up}
\label{eqn.psi}
\end{equation}

\noindent \textbf{Remarks}. Note that (\ref{eqn.psi}) represents the probability that an arbitrary in-connection of $p_i$
is active, whereas (\ref{eqn.eta}) represents the probability that an arbitrary out-connection of $p_j$ is
active. (\ref{eqn.psi}) is simply an approximation to the system, 
while (\ref{eqn.eta}) follows immediately from the random choking strategy. 

Let $\bar{U}$ denote the expected (steady-state) per-connection upload rate averaged
over all peers in the system. Then,
\begin{equation}
\bar{U} = \bigg( \sum_{j=1}^{B-1} \bar{N}_j \bar{U}_j \bigg) / \bar{N}_{up}
\end{equation}
Let $W_i$ denote the the number of active in-connections of $p_i$. 
If $W_i=w$, then $p_i$ will transit from queue $i$ to queue $i+w$. 
Thus, we have $i+1\le i+w \le B$ and $w \le \bar{N}_{up}-1$.
Let $\tilde{W}_i$ denote the upper bound on $W_i$. It is easy to see that 
$\tilde{W}_i=\min(B-i, \sum_{\ell=0}^{B-1} \bar{N}_\ell -1) $. 
Therefore, $W_i$ follows a truncated binomial distribution with parameter 
$\big( \bar{N}_{up}, P(S_i) \big) $ with an upper bound given by $\tilde{W}_i$.
Thus, the probability that $w$ in-connections are active is given by 
\begin{equation}
P(W_i=w)=\binom{\bar{N}}{w} P(S_i)^w (1-P(S_i))^{\bar{N} - w} \bigg/ \sum_{\ell=1}^{\tilde{W}_i} P(W_i=\ell)
\end{equation}

Assuming that the $w$ connections start and complete piece transfers at the same time, 
$\gamma_{i,i+w}$ can be calculated as
\begin{equation}
\gamma_{i, i+w} =  \bigg( \bar{U} \cdot  P(W_i=w) \bigg) \bigg/ \sum_{\ell=1}^{\tilde{W}_i} P(W_i=\ell)
\label{eqn.gamma.i.w}
\end{equation}
Let $P(i,i+w)$ denote the probability that $p_i$ directly transit from queue $i$ to queue $j$ in one hop, i.e.,  
\begin{equation}
P(i,i+w)=P(W_i=w)
\end{equation}

\noindent \textbf{Remarks.} 
Note that our model assumes that a peer has the same expected \emph{per-connection} download rate when in different queues 
in steady state. However, different peers are allowed to have different departure rates when they are in different queues. For an arbitrary
peer in
$i$-th queue, its departure rate is given by  $ \sum_{\ell=i+1}^B
\gamma_{i \ell}$, and the time it spends in the $i$-th queue is
exponentially distributed with mean $1/ \sum_{\ell=i+1}^B  \gamma_{i
\ell}$.

\subsection{Download completion time}
Based on the analysis in Sections  \ref{sec.delta_t}, \ref{sec.uj},
\ref{sec.conn}, \ref{sec.transition.rates}, we can now solve
the set of equations given in (\ref{eqn:rate}) to find:
\begin{itemize}
\item the expected number of peers in each queue in steady state, 
\item the probability that a peer in $i$-th queue is concurrently downloading $w$ pieces, i.e., $P(i,i+w), \forall i$ and $1 \leq w \leq
B-i$, 
\item the expected transition rates $\gamma_{ij}$ with $i=0,1,...,
B-1$ and $j=i+1, ..., B$.
\end{itemize}
We can then compute the expected download completion time of a peer in steady state as follows.

Let $T_{i}$ denote the expected remaining download completion time of
a peer in the $i$-th queue. 
The expected time that a peer stays in the $i$-th queue is given by
$1/ \sum_{\ell=i+1}^B  \gamma_{i \ell}$. We can now compute $T_i$ recursively as follows:
\begin{eqnarray}
T_i &=& \big( 1/ \sum_{\ell=i+1}^B  \gamma_{i \ell} \big) + \sum_{\ell=i+1}^{B} P(i,\ell) \cdot T_{\ell} \\
&&\quad  \quad \quad \quad i=0, 1, ..., B-2 \nonumber \\
T_{B-1}&=& \big( 1/  \gamma_{B-1, B} \big)
\end{eqnarray}
Thus  $T_0$ gives us the expected  download completion time of a newly
arrived peer. $T_0$ can be efficiently computed using dynamic programming.

Note that a special case of this model occurs when a peer makes exactly $B$ transitions during the downloading process, i.e., visits each
queue
exactly once. This case has been considered in the simple model for
coalition choking strategy in our previous work
\cite{zhang11icnp}. 

\subsection{Model Validation}
We next validate our analytical model by comparing its numerical results with extensive simulations of a coalition using
the random choking strategy.  
Throughout our simulations, we model the arrival of  
peers to the swarm as a Poisson process with an arrival rate of $20$ peers/minute. 
Both the initial seed's and peers' upload capacities are set to $0.5$ pieces/second. 
The seed's duration of re-choking interval is set to $10$ seconds, and
we vary the re-choking interval from $10-30$ seconds. We consider a
file with $B=60$ pieces. 
Each simulation lasts a duration of $4000$ seconds, and we obtain data
from the steady state ($3000-4000$ seconds) 
when the number of peers in the system is stablized. 

\begin{figure}[htb!]
\centerline{
    \begin{minipage}{2.2in}
      \begin{center}
        \setlength{\epsfxsize}{2.2in}
\epsffile{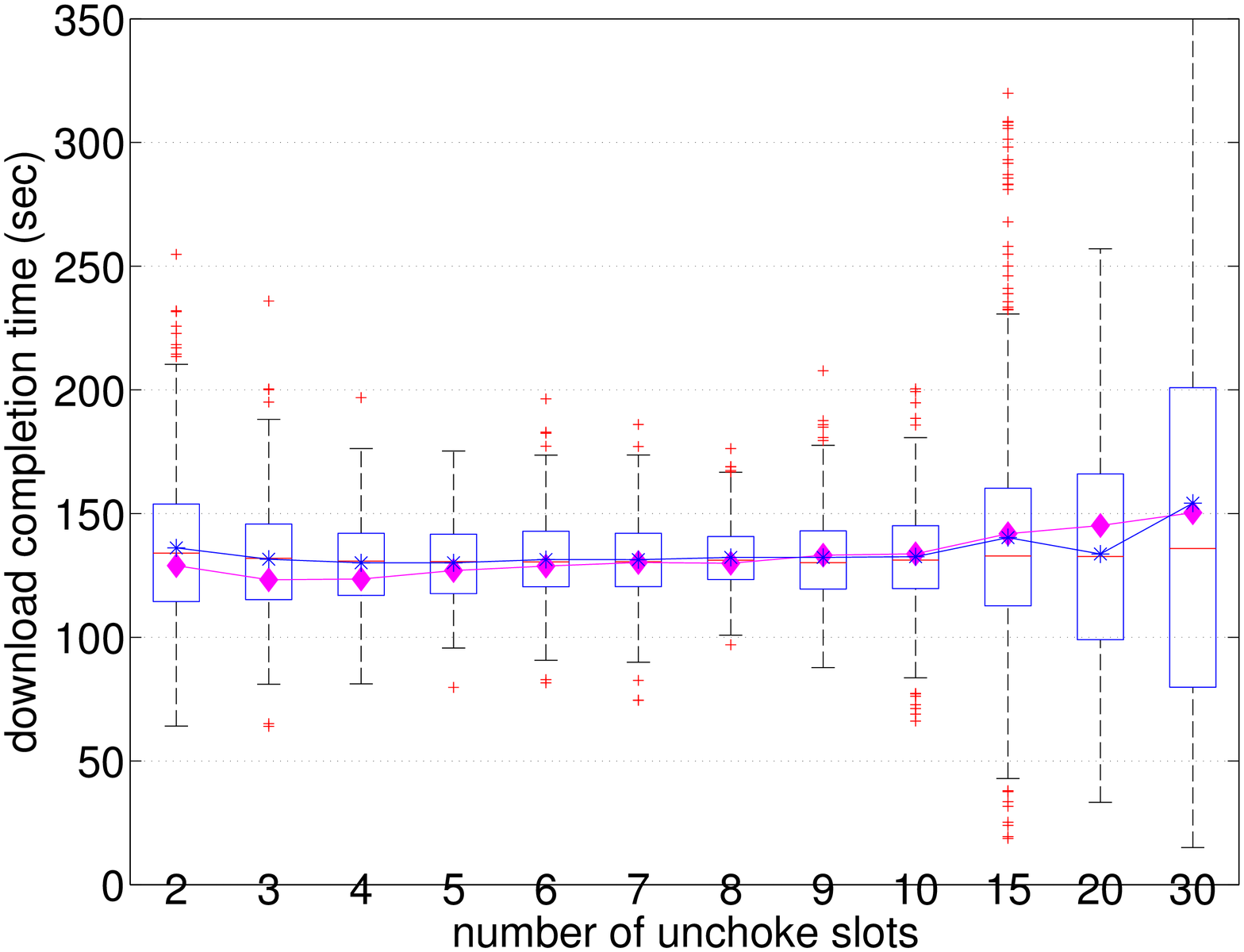}\\
       {}
      \end{center}
    \end{minipage}
}
\centerline{
    \begin{minipage}{2.2in}
      \begin{center}
        \setlength{\epsfxsize}{2.2in}
\epsffile{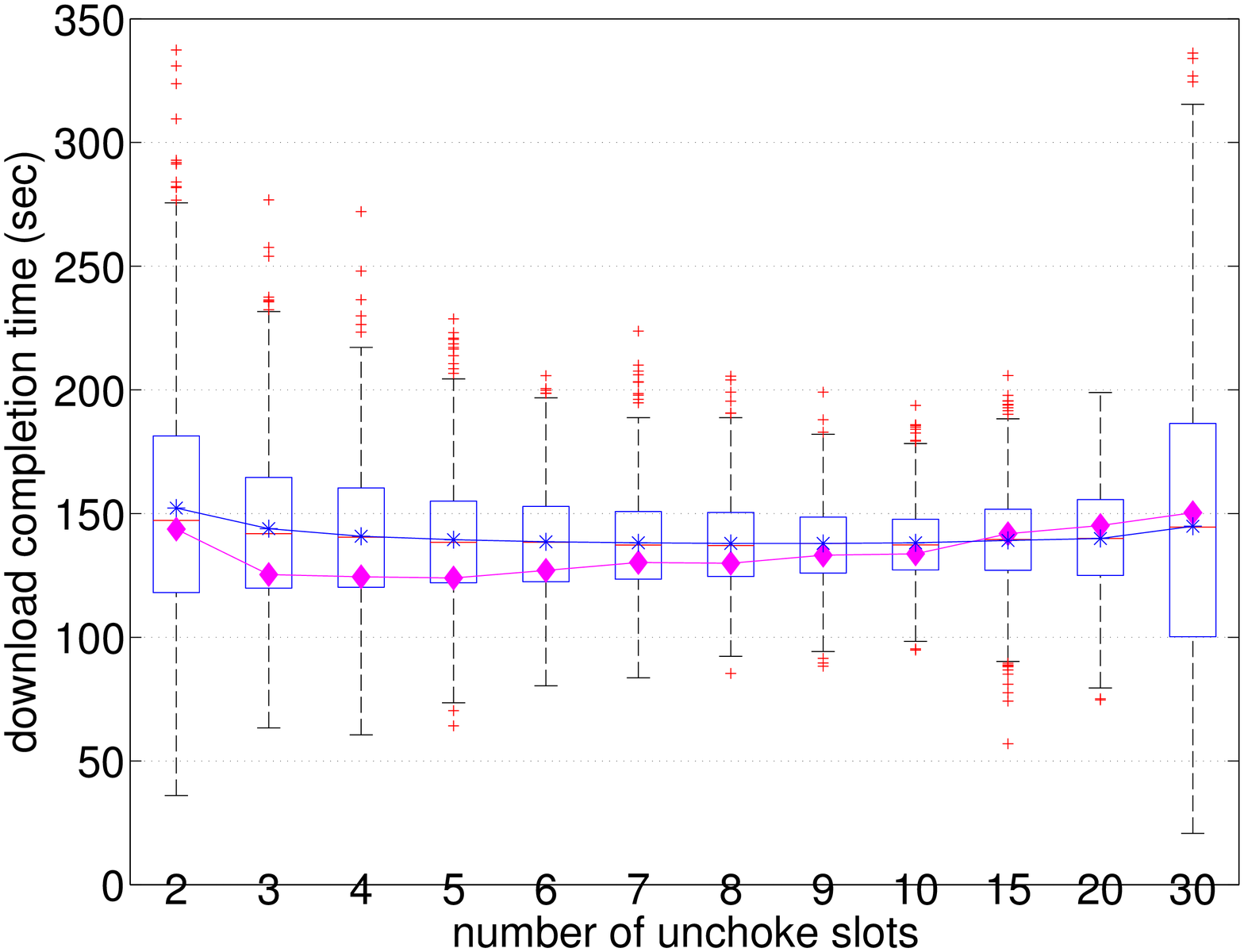}\\
       {}
      \end{center}
    \end{minipage}
}
 \caption{Comparison of download completion time between our model and
simulations with $B=60$ pieces. 
Top plot shows the case when $\delta_t=10$
seconds, and the second plot is for $\delta_t=20$ seconds. 
The boxplots show the simulation results, where the line
in each box represents the median, and the upper and lower edges of each box 
corresponds to the 25th and 75th percentiles respectively, and each '*' mark shows the average download completion time. The diamonds show
numerical results from our model. \label{fig:matlab_time_B_60}
}
\end{figure}

Figure \ref{fig:matlab_time_B_60} compares the numerical results
computed from our model with the simulation results. Figure
\ref{fig:matlab_time_B_60} shows that for a given value of $\delta_t$,
there indeed exists an optimal value of $k$, the number of unchoking slots. 
For example, when $\delta_t=10$ seconds, the optimal value of $k$ is around $4$.
We find that our model results in general matches well with the
simulation results.

\begin{figure}[htb!]
\centerline{
    \begin{minipage}{2in}
      \begin{center}
        \setlength{\epsfxsize}{2in}
\epsffile{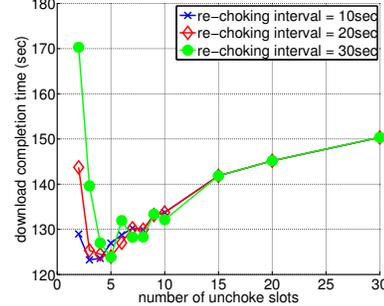}\\
       {}
      \end{center}
    \end{minipage}
}
 \caption{Download completion time predicted by our model for $B=60$.
\label{fig:matlab_time_B_60_intv_compare}
}
\end{figure}

Figure \ref{fig:matlab_time_B_60_intv_compare} shows the numerical
results of our model for different re-choking interval lengths. We see that 
when the number of unchoke slots is around 5, the expected download completion time is the lowest for $\delta_t=20, 30$ seconds, and close 
to the lowest for $\delta_t=10$ seconds. 
Note that in Figure
\ref{fig:matlab_time_B_60_intv_compare}, the lower bound for download completion time is $120$ seconds, as the file size is $B=60$ pieces
and $u_p=0.5$ piece/second.
When $k$ is small (less than $4$), the expected download completion time is shorter when $\delta_t=10$ seconds than $\delta_t=20, 30$
seconds.
This is because that longer re-choking interval can lead to decreased interestedness of a downloader in its data uploaders, i.e.,
$P(B_{ij})$ 
decreases with increasing $\delta_t$. 
However, as $k$ gets larger, the decreased interestedness can be
compensated for by an increased probability that a peer is unchoked by some
other peer, i.e., $P(A_{ij})$ increases with $k$. This is apparent in Figure \ref{fig:matlab_time_B_60_intv_compare} where the expected
completion times 
across different re-choking intervals approach each other when $k$ is large. In the extreme case when the number of unchoke
slots is very large (e.g., 30),
the re-choking interval has no impact on the system performance. This
is intuitive, as each peer is always unchoked by all other peers. 

\subsubsection{Implications of The Model}
Our analytical model illustrates that indeed there exist optimal
values for $\delta_t$ and $k$ when peers adopt the random choking strategy. Note that our
model does not try to find an optimal within-coalition choking
strategy for a coalition. Instead, our model
demonstrates that even the simple random choking strategy can yield near optimal performance. 
This is quite appealing, as the simple random
choking strategy can be easily implemented in a distributed way at
virtually no cost. A system designer can simply use our
model to find the optimal values of $\delta_t$ and
$k$ for given values of file size, and peers' arrival rate. 

\subsection{Extensions of The Model}
We note that our model can be generalized to other BitTorrent-like swarming system. For example, in our previous work
\cite{zhang11icnp}, we find $P(S_{ij})$ (probability of the connection from $p_j$ to $p_i$ being active) and $\bar{U}_j$
(per-connection upload rate from $p_j$) for a BitTorrent swarm. If we plug these equations into our model, we can analyze the performance
of a BitTorrent swarm. Second, for a coalition of peers with heterogeneous upload capacities, we can simply use the average capacity over
all the peers in the coalition as the $u_p$ in our model. A more
correct analysis would involve dividing the coalition into multiple
capacity classes, each class consisting of peers with approximately
the same upload capacities, and then treat each queue as a system of multiple
queues, each corresponding to a capacity class. We can then apply
the analytical machinery described in this section to analyze the system. Nevertheless, 
our simulations in Section \ref{sec.form.impact} show that a coalition with
heterogeneous peers can also significantly improve performance.

\section{Piece Selection Strategy Improves Data Availability of Coalition} \label{sec.replication}

\subsection{Strategy Design}
In this section, we propose a piece selection strategy (for data
replication in a coalition) to improve the data availability within a
coalition. This strategy is referred to as \textit{Peer-Balance Rarest-First Piece Selection} strategy.
This strategy specifies that,
a peer in a coalition first finds the set $S$ of pieces which are the rarest across the whole coalition (i.e., those pieces are possessed
by the least
number of peers in the coalition), when requesting pieces from other peers. 
Furthermore, when a peer finds that there is more than one piece in
$S$, then for each piece $k\in S$ and 
among all peers missing $k$, this peer identifies the peer with the
least number of pieces, denoted by $m_k$. 
Then, the peer selects a piece $b^*\in S$ to request for download,
where $b^*$ satisfies $m_{b^*}=\min_{k\in S} m_k$. Intuitively, 
this strategy gives the highest priority to the rarest piece (among
all pieces in the rarest set) which is not possessed by the peer
(denoted by $p_{b*}$) that possesses the smallest number of
pieces. Thus, $p_{b*}$ will be more likely to find an available
piece to download in future. In contrast, in BitTorrent's
rarest-first policy, a peer arbitrarily chooses a piece to request
from among all pieces in the rarest set.

We illustrate our strategy using the following example.  Table
\ref{tbl.piece} depicts matrix $V$ whose rows represent pieces of the
file, and the columns represent the peers. $v_{ij}$ (the $(i,j)$-th
entry of $V$) is set to 1 if peer $i$ has piece $j$, and 0 otherwise. 
Assume that peer $5$ is unchoked
by peer $6$. peer $5$ finds that the two rarest pieces are piece $1$
and $3$, both only replicated twice in the swarm. The peers that do
not have piece $1$ are peers $2, 3, 4,$ and $5$, each with $3, 2, 3,$ and
$2$ pieces respectively. Among these peers, the peer with the least
number of pieces are peers $3$ and $5$, each only in possession of $2$ pieces, i.e., $m_1=2$.
However, we find $m_3=1$, as peer $1$ misses piece $3$ (a piece in the
rarest set) but peer $1$ only owns one piece. If peer $5$ selects
piece $3$ to request for download (from peer $6$), then the poorest
peer (i.e., peer $1$) in the system now has one more
available piece that it can request in future. However, if peer $5$
requests for piece $1$, which peer $1$ is not interested
in, then the total number of pieces that are of interest to peer $1$
does not increase. Thus, our strategy is ``socialist'' with the
objective of helping the poorest peers first in order to increase the
overall data availability. 

\begin{table}[htb!]
\begin{center}
\begin{small}
\begin{tabular}{|c|c|c|c|c|c|c|}
\hline
 &peer 1 & peer 2 &peer 3 &peer 4 &peer 5 &peer 6  \\
\hline
piece 1  &1 & 0 & 0 & 0 & 0 & 1  \\
 piece 2 & 0 & 1 & 1 & 1 & 0 & 0  \\
 piece 3& 0 & 0 & 1 & 0 & 0 & 1  \\
 piece 4& 0 & 1 & 0 & 1 & 1 & 0  \\
 piece 5& 0 & 1 & 0 & 1 & 1 & 0  \\
\hline
\end{tabular}
\end{small}
\end{center}
\caption{Matrix $V$ for the example for piece selection strategy.}\label{tbl.piece}
\end{table}

\subsection{Data Availability} 
We next show how our piece selection strategy helps improve the data
availability in a coalition. 
Consider a flash crowd scenario (e.g., in data publishing phase) 
in which peers in a coalition request data
from an external source  before all distinct pieces have been disseminated into the coalition. 
Initially, there is no data piece in the coalition and
peers request and download data from the seed. Once a peer
receives a complete piece, it shares it with other peers in the
coalition. Note that only the seed can provide a new piece
to the coalition over time. Thus, in terms of the cumulative number of
distinct pieces that can be received by the coalition, the best
scenario occurs when the seed's upload capacity is entirely used 
to upload distinct pieces to the coalition. Let $t^*$ denote the time 
when the coalition receives all distinct pieces in this scenario, and
let $N_c(t)$ denote the number of distinct pieces
received by peers in a coalition at time $t$ in the best scenario,
with $t\in [0, t^*]$. Note that $N_c(t)$ serves as an upper bound for 
available distinct pieces in the coalition at time $t$. 

Since our data replication strategy always selects a piece that is
missing across the whole coalition, our strategy can achieve
$N_c(t)$. However, it is interesting to note that the rarest-first
strategy (even globally within the coalition) cannot achieve
this upper bound $N_c(t)$. To quantify this difference,
we introduce a metric referred to as availability loss. 

Let $N_d(t)$ denote the number of distinct pieces in the coalition at time $t$ when rarest-first strategy is adopted by peers in the
coalition. We define \textit{availability loss}, denoted by $L(t)$, as follows.
\begin{displaymath}
L(t) = (N_c(t) - N_d(t)) /N_c(t), \quad t\in [0, t^*]
\end{displaymath}
And define \textit{average availability loss} as 
\begin{displaymath}
\bar{L} = \int_0^{t^*} L(t^*) / t^*
\end{displaymath}

We next use simulations to show that the rarest first strategy
can lead to a high average availability loss under certain conditions. 
Consider $40$ homogeneous peers (with same upload capacity) that start
downloading from an initial seed uniformly at random in the first
$20$ seconds of the simulation. The initial seed's upload capacity is $1$ piece per second (i.e., 0.025 piece per second per peer). 
The durations of both re-choking and optimistic unchoking intervals are set to $40$ seconds.
Since the number of peers is large, a peer $i$ will equally likely unchoke or
choke other peers during each re-choking interval. In other words, any
other peer is equally likely to appear in peer $i$'s unchoke set (which has 5 peers). 
Thus, on average, each peer is roughly unchoked by $5$ other peers
and its downloading rate equals to the upload capacity of a
single peer. 

\begin{figure}[htb]
\centerline{
    \begin{minipage}{1.7in}
      \begin{center}
        \setlength{\epsfxsize}{1.7in}
	  \epsffile{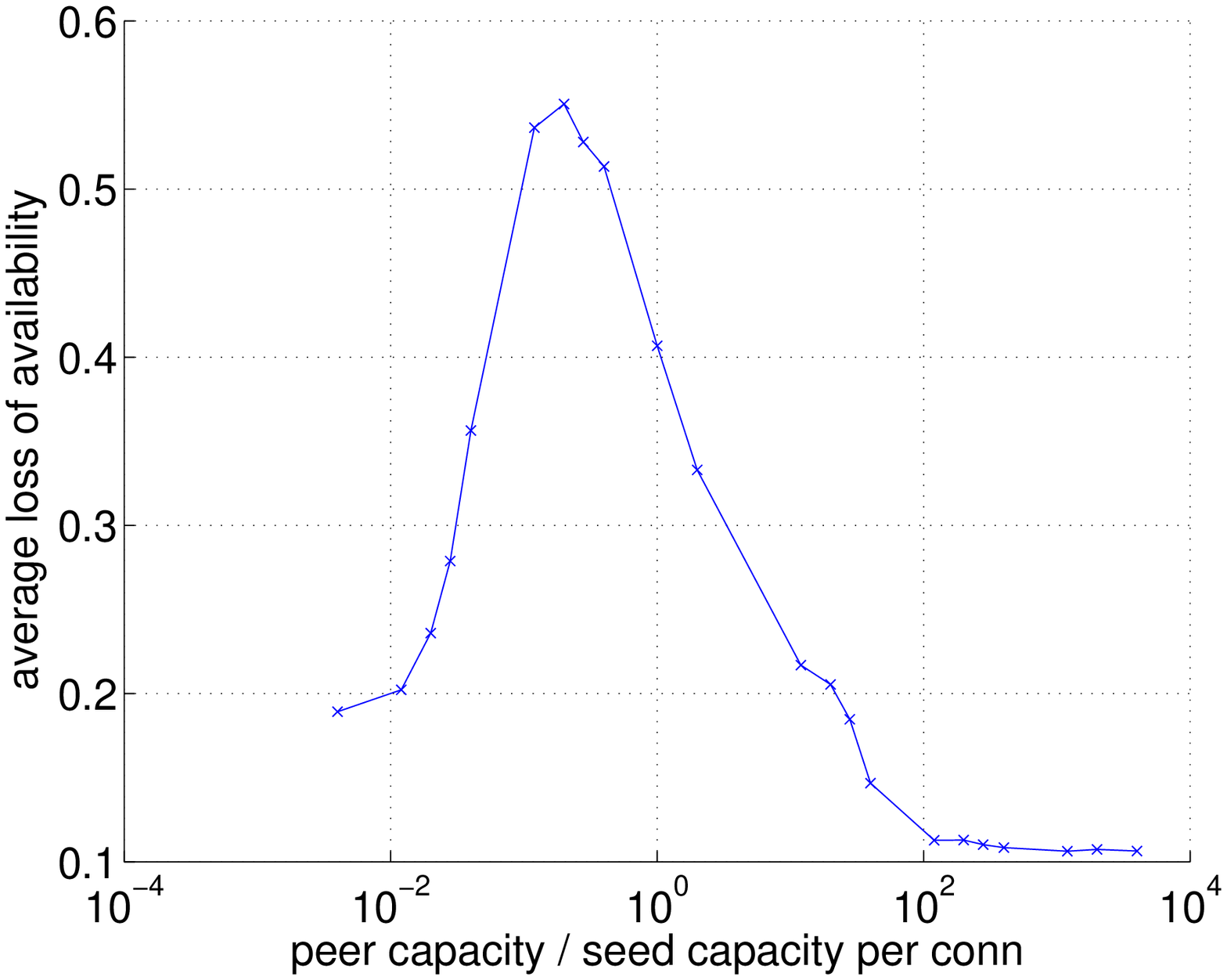}\\
       {}
        \caption{
        Average availability loss when 40 peers download 500 pieces from an initial seed in $1200$ seconds.
        }
        \label{fig:avail_loss_40peers}
      \end{center}
    \end{minipage}
    \begin{minipage}{1.7in}
      \begin{center}
        \setlength{\epsfxsize}{1.7in}
        \epsffile{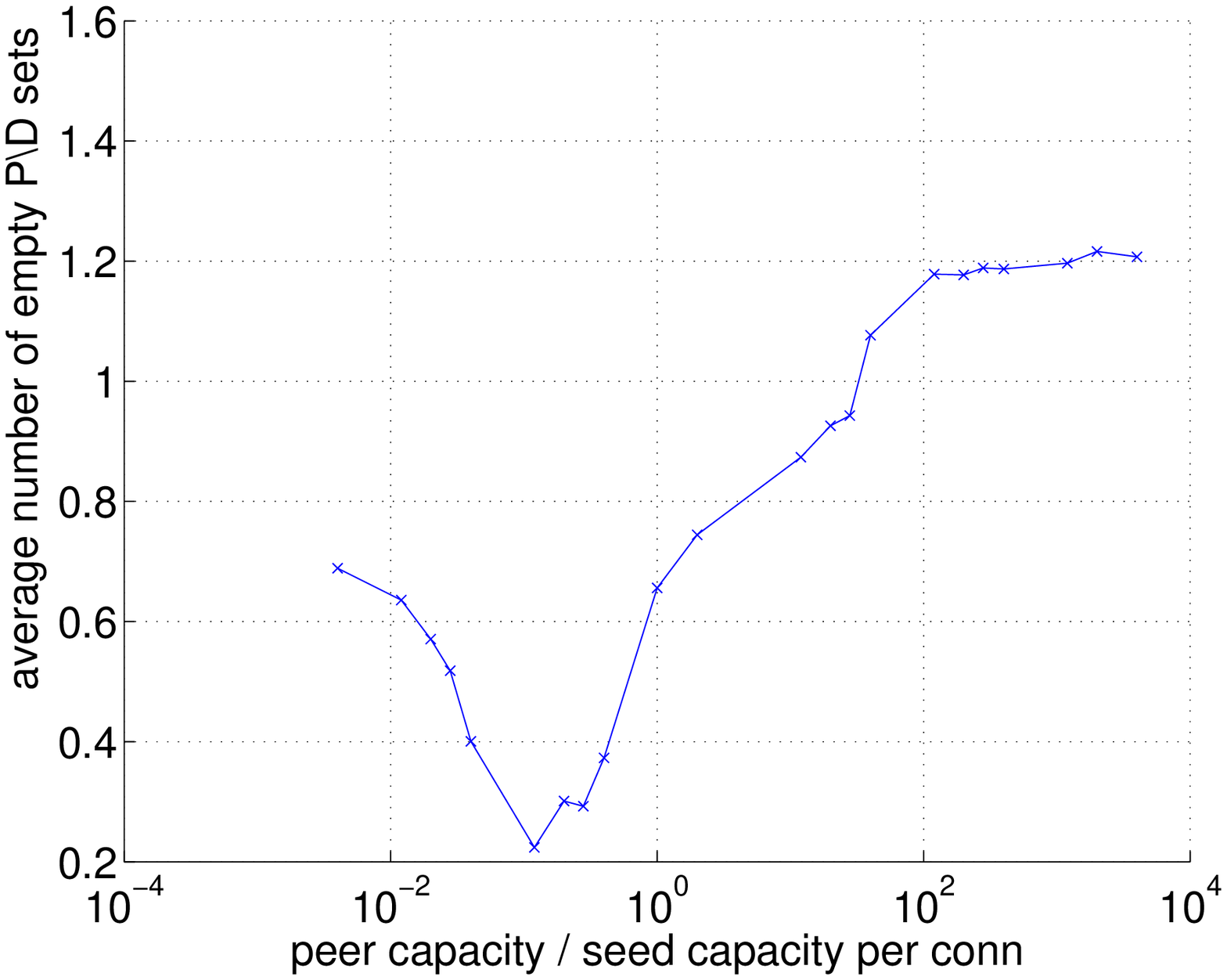}\\
       {}
        \caption{
        Average number of empty $P\setminus D$ sets.
        }
        \label{fig:pdset_40peers}
      \end{center}
    \end{minipage}
}
\end{figure}

We consider a global (i.e., within coalition) rarest first
algorithm. That is, a peer selects the rarest piece that has the least
copies among all $40$ peers in this swarm. 
All peers have the same upload capacity, ranging
from $100$ to $0.0001$ pieces per second.
For each upload capacity, we run 10 simulations with different random
seeds and calculate the average availability loss
(defined earlier).
As shown in Figure \ref{fig:avail_loss_40peers}, we observe that if
the peers' upload capacity is high, there is almost no availability
loss; on the other hand, if the peers' upload capacity is low, the availability loss is very low. But, when the peers' upload capacity
is of the same order as the seed's upload capacity, the availability
loss is large (more than $50$ percent).

The large availability loss observed appears counter-intuitive at the
first glance. However, it can be explained
by peers' piece selection behavior. In BitTorrent-like file sharing, a
peer always requests a piece (from the seed or other peers) that is in its
partially finished set and not being actively downloaded. Thus, at the time when a peer selects a piece to request from the seed, if
this peer still has a partially finished but not a currently
downloaded piece, then this peer does not choose a new piece that has not been
disseminated into the swarm, even though it employs the rarest-first policy. 

More specifically, let $P_i(t)$ denote the partial set containing pieces that are partially downloaded by peer $i$ at time $t$. Let $D_i(t)$
denote the set containing pieces that peer $i$ is
downloading at time $t$. Clearly $D_i(t) \subseteq P_i(t) $. Peer $i$
requests pieces in $P_i(t)\setminus D_i(t)$,
because it wants to finish partially
downloaded pieces first. Since the set $P(t) \setminus
D(t)$ of peers are often non-empty, peers will frequently
request the seed for pieces that the seed has already sent to other
peers in the coalition. The reason for the non-emptiness of set
$P(t)\setminus D(t)$ is due to the periodic choking and unchoking behavior.
Let $\overline{E}_{PD}$ denote the average number of empty $P\setminus D$ sets when peers select pieces to request from the
initial seed. We plot $\overline{E}_{PD}$ in Figure \ref{fig:pdset_40peers}, and find
that $\overline{E}_{PD}$ is negatively correlated with $\bar{L}$,
which is consistent with our conjecture. 
We show further evidence in support of our conjecture by comparing the number of times that peers have
non-empty sets $P_i(t)\setminus D_i(t)$ and the number of received distinct pieces every $10$ seconds. A cross-correlation analysis of the
two time series
shows that the maximum lag (with correlation coefficient $>0.9$) between them is $40$ seconds which is the average downloading time from the
seed by a single peer. Due to space limits, we present this result in a technical report.
Part of our ongoing research is
to understand why the peak availability loss occurs when ratio of peer
capacity to seed capacity attains a specific value.

\section{Impact and Stability of coalition} \label{sec.form.impact}
In this section, we investigate the impact of coalition on the performance of a swarm,
and unlike \cite{zhang11icnp}, we investigate whether stable coalitions
result in a swarm where peers have heterogeneous capacities. 
Further, our simulations here employ data sets
collected from real-world swarming systems \cite{piatek07bittyrant},
unlike \cite{zhang11icnp} which simulates a 
swarm with two capacity classes. 

\subsection{Impact of Coalition On Data Swarming Performance}
We now study the impact of coalitions on a population of peers with
heterogeneous upload capacities taken from real-world data swarms
\cite{piatek07bittyrant}. The distribution of upload capacities is 
shown in Figure \ref{fig:plot_capc_emprc_bw}.
The figure shows that majority of the peers in a swarm are low
capacity peers. The lowest capacity is $42.96$KBps. $90\%$ of 
peers have capacity below $391.45$KBps, while very few peers have
capacity more than $10$MBps. 

In our simulations, peers arrive according to a Poisson process
with an arrival rate of $20$ peers/minute. If a peer is not in
a coalition, the peer uses the regular BitTorrent algorithms with 
$\delta_t=10$sec and $k=5$ ($4$ slots for Tit-for-Tat and $1$ for optimistic
unchoke). There is one coalition in the system, and peers in the coalition use the
random choking strategy with $\delta_t=10$ seconds and $k=5$, the
optimal values as determined by our model. The upload capacities of peers are sampled from
the empirical distribution given in Figure \ref{fig:plot_capc_emprc_bw}.
The seed's upload capacity is set to $0.5$ pieces/second with
$\delta_t=10$ seconds. The shared file has $B=80$ pieces.   
In our simulations, individual peers significantly improve their performance by
joining the coalition compared to the case when they do not. These results are not reported here due to space
limits and we choose to focus on the swarm performance instead. 
\begin{figure}[htb!]
\centerline{
    \begin{minipage}{1.8in}
      \begin{center}
        \setlength{\epsfxsize}{1.8in}
\epsffile{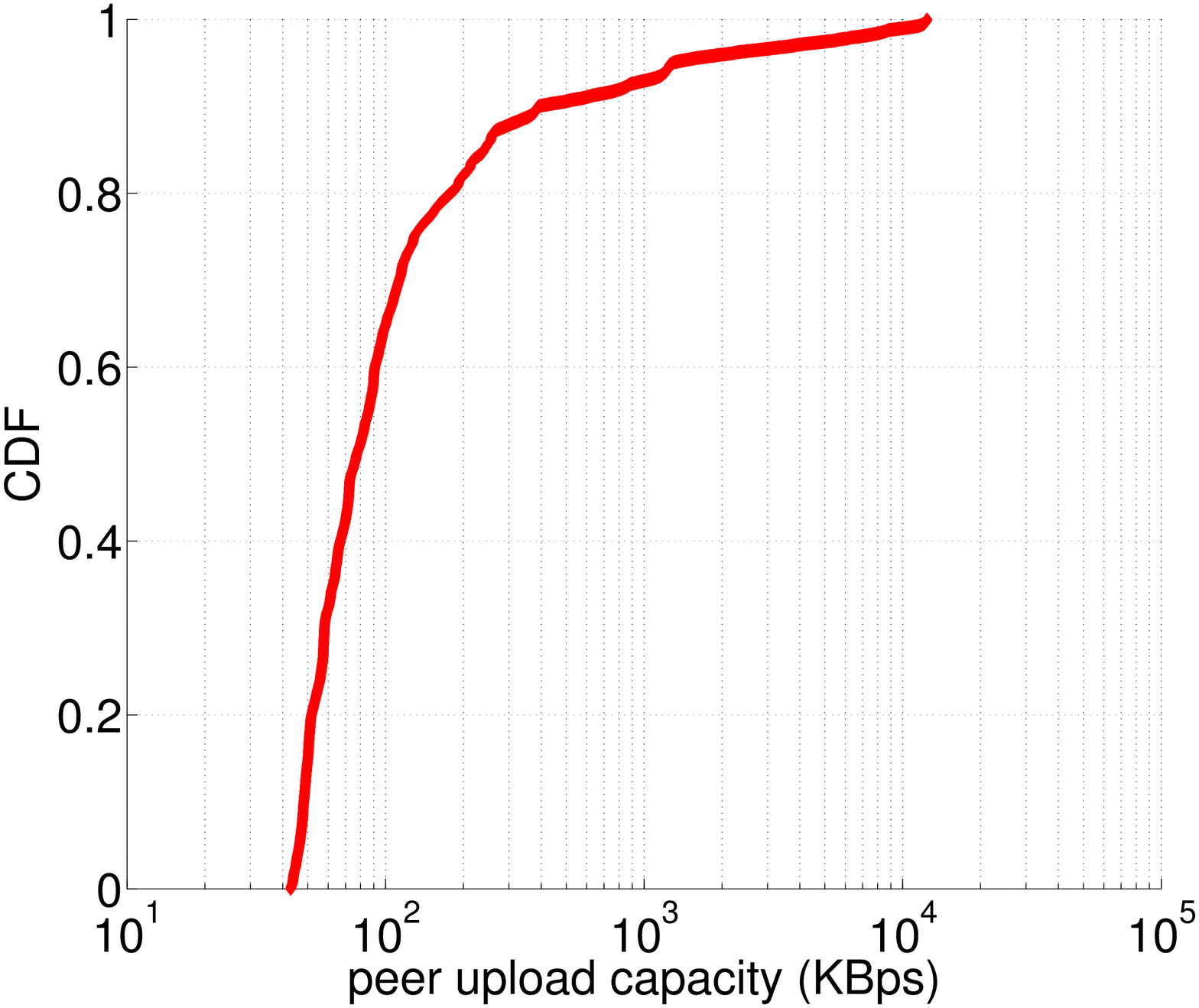}\\
       {}
      \end{center}
\label{fig:plot_capc_emprc_bw}
    \end{minipage}
}\caption{Empirical upload capacity distribution from \cite{piatek07bittyrant}. \label{fig:plot_capc_emprc_bw}}
\end{figure}

\subsubsection{Coalition formed by low capacity peers}
We first investigate whether a coalition formed by low capacity peers improves performance.
We let peers with capacity below the $50$-th percentile (i.e., in the
range  $[42.96, 77.40]$KBps) in Figure \ref{fig:plot_capc_emprc_bw} form
a coalition. We also consider the $70$-th and $90$-th percentiles,
corresponding to capacities about $112.87$ KBps and $391.45$ KBps respectively, for forming coalitions. 

In Figure \ref{fig:plot_capc_emprc_bw_static_coal_B_80_boxplot_lowcapc}, 
we plot the boxplots of the steady state download completion times for
different coalition sets, and compare them with the case with no coalition. 
We can see that coalition significantly improves the overall
performance of the swarm. For example, if peers with capacity below 
the $90$-th percentile join the coalition, the 
average download completion time is reduced by over $20\%$ compared to
the case when there is no coalition. In addition, the variance of download completion times of the
whole swarm is also significantly reduced when peers with capacity
below the $70$-th or $90$-th percentiles form a coalition.

\subsubsection{Coalition of randomly chosen peers}
\begin{figure}[htb!]
\centerline{
    \begin{minipage}{1.7in}
      \begin{center}
        \setlength{\epsfxsize}{1.7in}
\epsffile{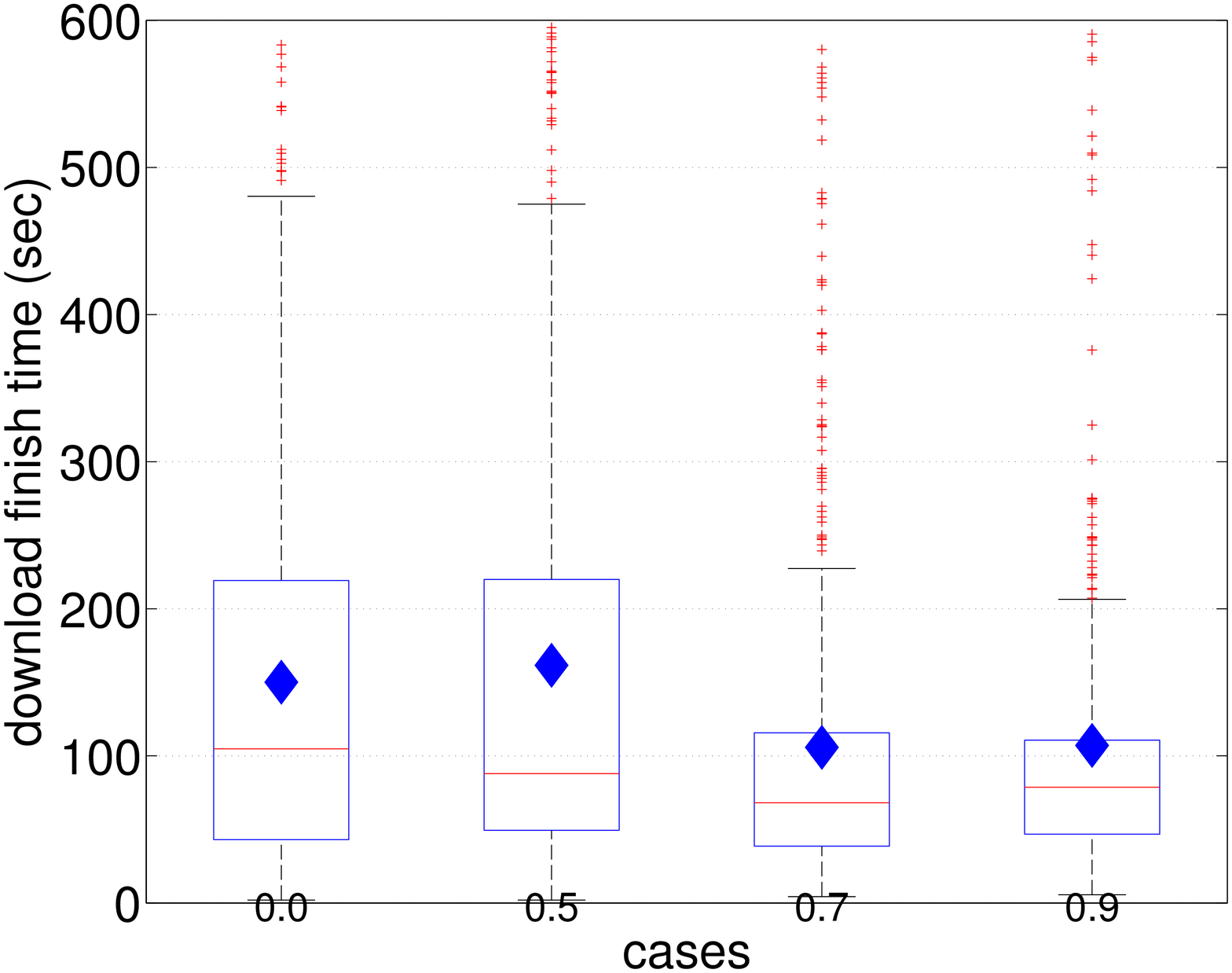}\\
       {}
      \end{center}
\caption{
Boxplots show the median and percentiles ($25, 75$). Diamonds show the average.
Four cases: no coalition, and coalitions of users with capacity below the
$50$-th, the $70$-th and the $90$-th percentiles. 
\label{fig:plot_capc_emprc_bw_static_coal_B_80_boxplot_lowcapc} }
    \end{minipage}
    \begin{minipage}{1.7in}
      \begin{center}
        \setlength{\epsfxsize}{1.7in}
\epsffile{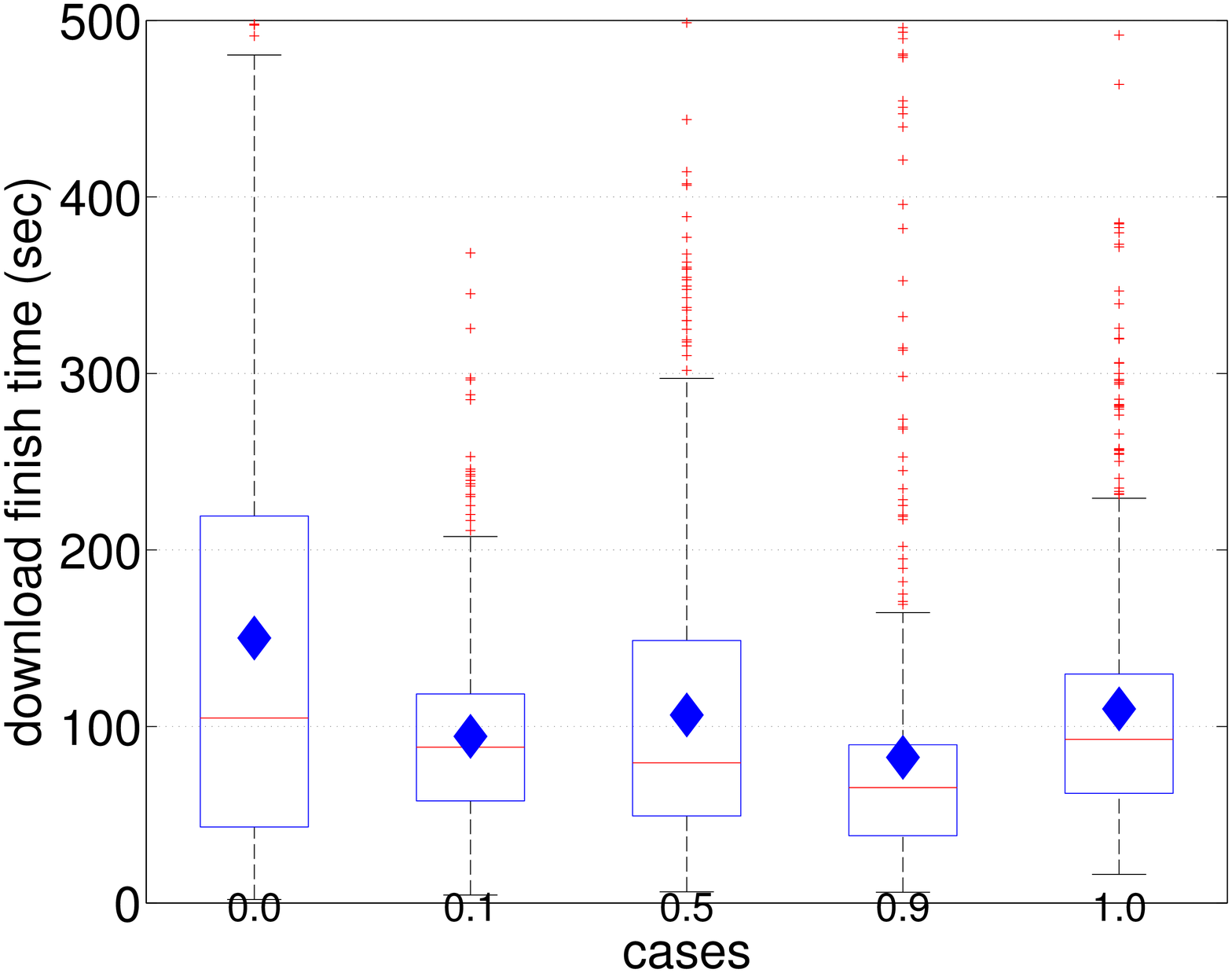}\\
       {}
      \end{center}
\caption{Boxplots show the median and percentiles ($25, 75$), and diamonds show the average download completion time. The labels on x-axis
represents
different $p_{join}$.
\label{fig:plot_capc_emprc_bw_static_coal_B_80_boxplot} }
    \end{minipage}
 }
\end{figure}

A newly arrived peer joins the coalition with probability $p_{join}$. We vary $p_{join}$ as $\{0, 0.1, 0.5, 0.9, 1.0\}$, and for each value
of
$p_{join}$, we record the file download completion time of peers that
join and complete their downloads during steady state. 
In Figure \ref{fig:plot_capc_emprc_bw_static_coal_B_80_boxplot}, we
plot the boxplots of the steady state download completion times. 
We see that the overall system performance is significantly improved 
even when peers randomly join the coalition. 
It is interesting to note that the best performance occurs when $90\%$
of peers join the coalition, instead of $100\%$. This can be explained
as follows. When all peers join the coalition, the coalition
has more low capacity peers, which leads to a worse
performance than the case when $p_{join}=0.9$. Even then, it 
significantly outperforms the case with no coalition. 
Note that the extra $10\%$ ($=100\%-90\%$) of the peers are mainly low
capacity peers, as can be seen from Figure \ref{fig:plot_capc_emprc_bw}.

\noindent \textbf{Remarks.} If we compare the fourth boxplot (from the
left) in Figure \ref{fig:plot_capc_emprc_bw_static_coal_B_80_boxplot} with the boxplots in Figure
\ref{fig:plot_capc_emprc_bw_static_coal_B_80_boxplot_lowcapc}, we see
that a coalition of randomly chosen peers yields better
overall swarm performance than the coalition composed only of low
capacity peers. This is because a few of the high capacity peers join the
coalition when peers are randomly picked to join the coalition. 

We also investigate the case when there are multiple coalitions in the swarm. 
Specifically, we simulate two coalitions in a swarm with peer
capacities chosen according to the empirical distribution in \cite{piatek07bittyrant}. 
A newly arriving peer with capacity below the $q_{low}$-th percentile of the distribution will join coalition 1, but if its capacity is
above
the $q_{high}$-th percentile of the distribution, it will join coalition 2.
We vary  $q_{low}$ as $\{10, 50, 90\}$ when $q_{high}=10$, and vary $q_{low}$ as $\{10, 50,
70\}$ when $q_{high}=30$. Again, we find that in each scenario,
forming two coalitions significantly improves the whole swarm's performance. 

\subsection{Dynamic Coalition Formation}
Similar to \cite{zhang11icnp}, we next study whether coalition size
converges to a fixed value when peers use the cooperation-aware
better-response strategy proposed in our
previous work \cite{zhang11icnp}. We improve the strategy in
\cite{zhang11icnp} to allow a peer to join any coalition regardless of its
capacity class.  

\subsubsection{Improved cooperation-aware better-response strategy} 
Suppose that there are multiple coalitions in the swarm.
The basic idea of this strategy is that a peer always attempts to join the coalition
with maximum rate, but will not change membership if its own rate is
no less than its coalition average rate (discounted by a non-cooperation factor, denoted
as $\beta$). The larger the $\beta$, the more non-cooperative a peer is. 
Specifically, peer $j$ makes a decision every $r\cdot \delta$ time units, where
$\delta$ is the re-choking interval length, and $r>0$ is referred to as 
the {\em patience factor}. A larger $r$ implies that the peer makes
decisions of whether to remain in the coalition or not over a longer time
period. When making a decision, the peer compares 
its own download rate with the average download rate of its own
coalition (if it is in a coalition), and the maximum average download
rate across all coalitions. All three rates are averaged over the last
$r\cdot \delta$ time units. The average download rate of a coalition is
discounted by $\beta$. If a peer is currently not in any coalition and its own rate is less than the
maximum rate, then it joins the coalition with the maximum rate. If a
peer is currently in a coalition and its own rate is no less than its
coalition's average rate, then it stays in the coalition; otherwise if
its own rate is less than its coalition rate and less than the maximum
rate, then it joins the coalition with maximum rate (if not in it
yet). A peer leaves its coalition if its rate is less than its
coalition's rate and its coalition is the one with maximum rate. 
In the following, we report our results on a swarm consisting of
only one coalition and on a swarm with two coalitions. 

\subsubsection{Only one coalition in a swarm} 
\begin{figure}[htb!]
\centerline{
    \begin{minipage}{1.7in}
      \begin{center}
        \setlength{\epsfxsize}{1.7in}
\epsffile{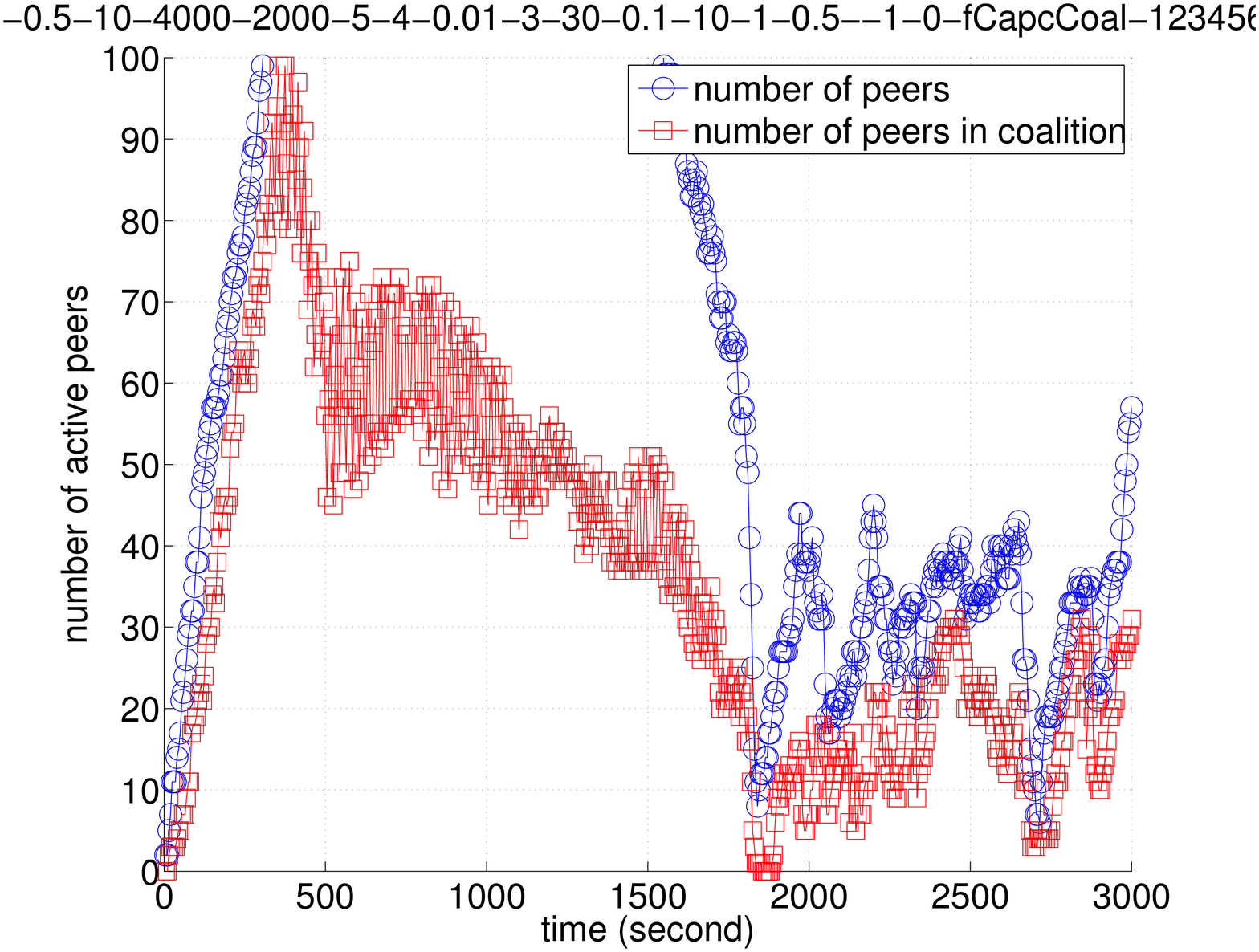}\\
       {}
      \end{center}
\caption{Number of peers in a swarm with one dynamic coalition.
\label{fig:num_peers_emprc_bw_startP_0.1_varyIntvFctr_10_nonCoop_0.5} }
    \end{minipage}
    \begin{minipage}{1.7in}
      \begin{center}
        \setlength{\epsfxsize}{1.7in}
\epsffile{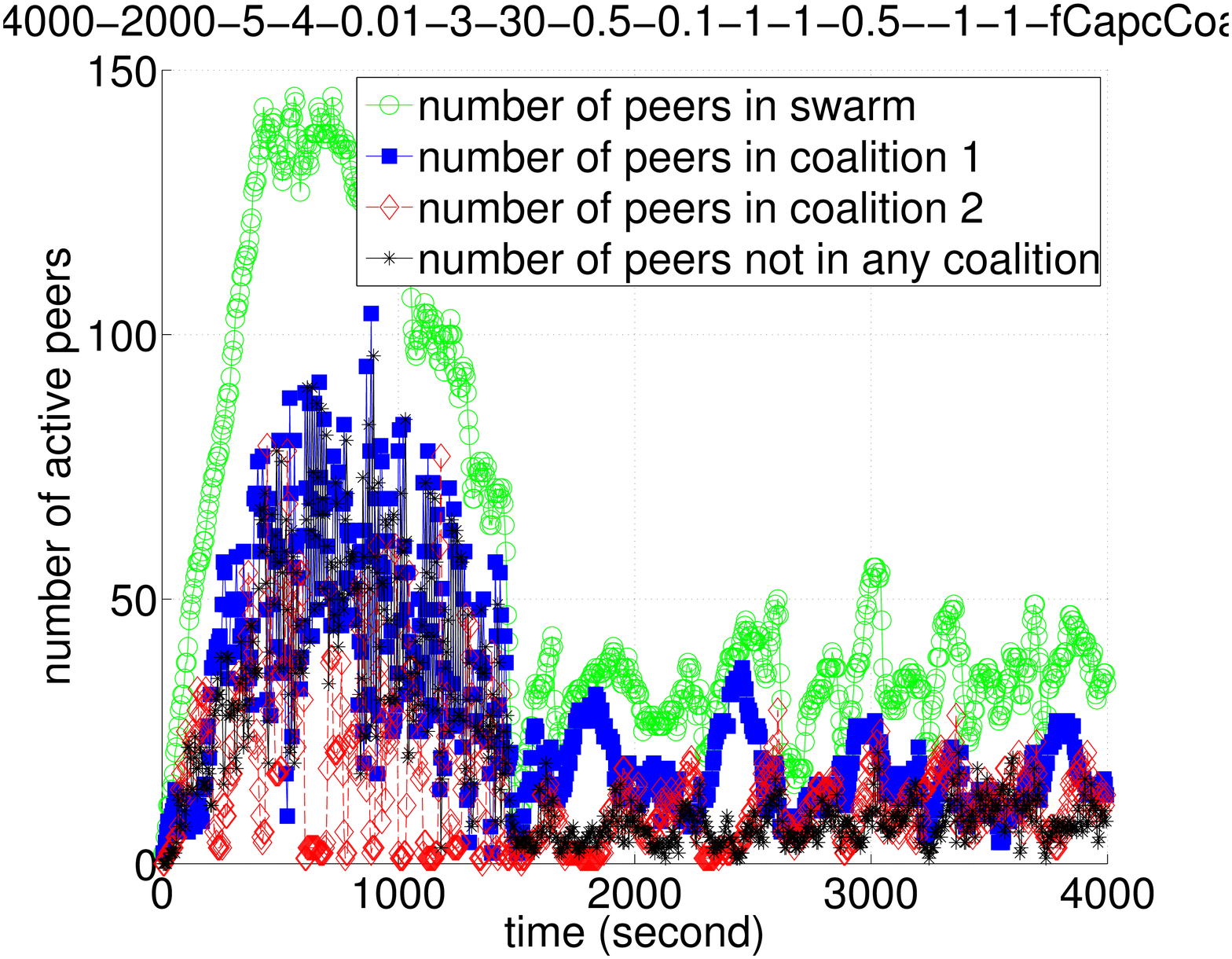}\\
       {}
      \end{center}
\caption{Numbers of peers in a swarm with two dynamic coalitions.
\label{fig:num_peers_dynamic_emprc_bw_coal_1_coal_2startP_0.5_varyIntvFctr_1_nonCoop_0.5_coal_0} }
    \end{minipage}
 }
\end{figure}

In this scenario, a newly arrived peer joins the coalition with a fixed probability $q_{init}$, taking values
$0.1, 0.5, 0.9$. Peers update their
coalition membership three re-choking intervals after their arrival, using the above mentioned  cooperation-aware better-response strategy.
We consider different
values of $r$ and $\beta$, viz, $r=1, 5, $ or $10$, and $\beta=0.1, 0.5, $ or $1.0$.
In our simulations, the coalition size is always stable after $2000$
seconds into the simulation. For example, 
Figure
\ref{fig:num_peers_emprc_bw_startP_0.1_varyIntvFctr_10_nonCoop_0.5}
shows that the number of peers in coalition set is stable in
steady state when $r=10, \beta=0.5, q_{init}=0.1$. Even though a newly arrived peer joins the coalition with low probability $0.1$,
eventually the 
coalition is attractive enough such that more than $50\%$ of the peers
end up in a coalition on average. \cite{zhang11icnp} shows the dynamic
stability of coalitions, but each coalition only consists of peers
with the same capacity. Our results here show that the coalition size
stabilizes  in steady state, even if the coalition consists of peers with heterogeneous capacities. 

We compare the average fractions of peers that are in the coalition in steady
state across all different combinations of $\beta, r$, and $q_{init}$. 
Similar to the two capacity class scenario considered in
\cite{zhang11icnp}, we observe that as $\beta$ increases,
the coalition sizes decreases, as peers become less cooperative. Also
the coalition size increases as $r$ increases (peers are more
patient). And, when $\beta<0.5$, regardless of values of $q_{init}$ and $r$, the coalition is always able to keep more than $50\%$ peers
(out of the whole swarm) in the coalition.

\subsubsection{Two coalitions in a swarm} 
In a swarm with two coalitions, we find that our 
cooperation-aware better response strategy yields
stable coalitions in steady state for any combination of $\beta$ and
$r$, with values in $\{0.1, 0.5, 1.0\}$ and $\{1, 5,
10\}$ respectively. 
We simulate three cases, viz. a peer
with capacity below the $x$-th ($x = 50,70,90$) percentile initially
joins coalition 1. We observe stable coalitions in each scenario. 
Figure \ref{fig:num_peers_dynamic_emprc_bw_coal_1_coal_2startP_0.5_varyIntvFctr_1_nonCoop_0.5_coal_0} shows the number of peers in 
swarm when $\beta=0.5$ and $r=1$.

\section{Conclusion and Future Work}\label{sec.con}
Based on the encouraging preliminary results in our previous work \cite{zhang11icnp}, 
we proposed a detailed yet flexible analytical model for coalitions in
a swarming system. Our model yields optimal parameter settings for the
random choking strategy for a coalition.  We also proposed a
piece selection strategy to improve the data availability for a
coalition. We demonstrated that coalitions yield significant
performance improvement for the whole swarm, and we further
showed that coalitions can be reached and remain stable if peers adopt
an enhanced cooperation-aware better-response strategy when they
dynamically join or leave a coalition.  For future work, 
we plan to investigate the upload capacity allocation strategy of peers in a
coalition, and formally analyze the dynamic stability of coalitions.


\section{Appendix: Find $P(B)$} \label{sec.pb}

\noindent \textbf{Case where $i \ge j$.}
If $\ell>j$, we always have $P(B^\ell_{ij}=1)=0$. 
We also assume that 
$P(B_{ij}^1=1)=1-  \binom{B-j}{i-j} \big/  \binom{B}{i}$. 
 
In $\ell$-th slot with $2 \le \ell \le \varphi_j $ and $\ell <=j $,
\begin{eqnarray}
P(B_{ij}^\ell=1) &=&  P(B_{ij}^{\ell-1}=1)  P(B_{ij}^\ell=1 | B_{ij}^{\ell-1}=1) \nonumber \\
&& + P(B_{ij}^{\ell-1}=0)  P(B_{ij}^\ell=1 | B_{ij}^{\ell-1}=0) \nonumber \\
 &=&  P(B_{ij}^{\ell-1}=1)  P(B_{ij}^\ell=1 | B_{ij}^{\ell-1}=1) \nonumber 
\end{eqnarray}
If $B_{ij}^{\ell-1}=0$, then $p_i$ is not interested in $p_j$ in $\ell$-th slot.

\begin{eqnarray}
&& P(B_{ij}^\ell=1 | B_{ij}^{\ell-1}=1) \nonumber \\
&=& 1- P(B_{ij}^\ell=0 | B_{ij}^{\ell-1}=1)  \nonumber  \\
&=&1-  \frac{1}{P(B_{ij}^{\ell-1}=1)} \cdot 
\frac{ \binom{B}{\ell-1} \binom{B- (\ell-1)}{j-(\ell-1)} \binom{B-j}{i-(j-(\ell-1))}   }{   \binom{B}{i} \binom{B}{j} }
\label{eqn.neg.B}
\end{eqnarray}
 

\noindent \textbf{Case where $i < j$.}
If $\ell>j$, we always have $P(B^\ell_{ij}=1)=0$. 

\noindent In the $\ell$-th slot where $1\le \ell \le \varphi_j$ and $(\ell-1) < j-i $  and $\ell\le j$,
we have $P(B_{ij}^\ell=1)=1$.
 At the beginning of this slot, $p_i$ has $i+\ell-1 (<j)$ pieces, thus, $p_i$ is always interested in $p_j$.

In the $\ell$-th slot where $1\le \ell \le \varphi_j$ and $(\ell-1) \ge j-i$  and $\ell\le j$.
These are the cases that the probability of $p_i$ being interested in $p_j$ is less than one.
At the beginning of this slot, $p_i$ has $i+\ell-1 (>j)$ pieces, thus, $p_i$ has more pieces than $p_j$, but they 
have $\ell-1$ common pieces.

\begin{eqnarray}
P(B^\ell_{ij}=1) &=& P(B^\ell_{ij}=1 | B^{\ell-1}_{ij}=1 ) P ( B^{\ell-1}_{ij}=1) \nonumber 
\end{eqnarray}
 
\begin{eqnarray}
&& P(B^\ell_{ij}=1 | B^{\ell-1}_{ij}=1 ) \nonumber \\
&=& 1- P(B^\ell_{ij}=0 | B^{\ell-1}_{ij}=1 )\nonumber \\
&=& 1- \frac{1}{ P ( B^{\ell-1}_{ij}=1)} 
\frac{ \binom{B}{\ell-1} \binom{B- (\ell-1)}{j-(\ell-1)} \binom{B-j}{i-(j-(\ell-1))}   }{   \binom{B}{i} \binom{B}{j} }
\end{eqnarray}

Here, since $\ell-1 \ge j-i$, thus, $i-(j-(\ell-1))>=0$.


\end{document}